\documentstyle[twoside]{article}

\catcode`\@=11
\long\def\@makefntext#1{
\protect\noindent \hbox to 3.2pt {\hskip-.9pt  
$^{{\eightrm\@thefnmark}}$\hfil}#1\hfill}		

\def\@makefnmark{\hbox to 0pt{$^{\@thefnmark}$\hss}}	
	
\def\ps@myheadings{\let\@mkboth\@gobbletwo
\def\@oddhead{\hbox{}
\rightmark\hfil\eightrm\thepage}   
\def\@oddfoot{}\def\@evenhead{\eightrm\thepage\hfil
\leftmark\hbox{}}\def\@evenfoot{}
\def\sectionmark##1{}\def\subsectionmark##1{}}

\oddsidemargin=\evensidemargin
\newcounter{sectionc}\newcounter{subsectionc}\newcounter{subsubsectionc}
\renewcommand{\section}[1] {\vspace{12pt}\addtocounter{sectionc}{1} 
\setcounter{subsectionc}{0}\setcounter{subsubsectionc}{0}\noindent 
	{\tenbf\thesectionc. #1}\par\vspace{5pt}}
\renewcommand{\subsection}[1] {\vspace{12pt}\addtocounter{subsectionc}{1} 
	\setcounter{subsubsectionc}{0}\noindent 
	{\bf\thesectionc.\thesubsectionc. {\kern1pt \bfit #1}}\par\vspace{5pt}}
\renewcommand{\subsubsection}[1] {\vspace{12pt}\addtocounter{subsubsectionc}{1}
	\noindent{\tenrm\thesectionc.\thesubsectionc.\thesubsubsectionc.
	{\kern1pt \tenit #1}}\par\vspace{5pt}}
\newcommand{\nonumsection}[1] {\vspace{12pt}\noindent{\tenbf #1}
	\par\vspace{5pt}}

\newcommand{\textlineskip}{\baselineskip=13pt}

\def\abstracts#1#2#3{{
	\centering{\begin{minipage}{4.5in}\baselineskip=10pt\footnotesize
	\parindent=0pt #1\par 
	\parindent=15pt #2\par
	\parindent=15pt #3
	\end{minipage}}\par}} 

\renewenvironment{thebibliography}[1]
	{\frenchspacing
	 \ninerm\baselineskip=11pt
	 \begin{list}{\arabic{enumi}.}
        {\usecounter{enumi}\setlength{\parsep}{0pt}     
	 \setlength{\leftmargin 12.7pt}{\rightmargin 0pt} 
         \setlength{\itemsep}{0pt} \settowidth
	{\labelwidth}{#1.}\sloppy}}{\end{list}}

\newcounter{itemlistc}
\newcounter{romanlistc}
\newcounter{alphlistc}
\newcounter{arabiclistc}

\def\@citex[#1]#2{\if@filesw\immediate\write\@auxout
	{\string\citation{#2}}\fi
\def\@citea{}\@cite{\@for\@citeb:=#2\do
	{\@citea\def\@citea{,}\@ifundefined
	{b@\@citeb}{{\bf ?}\@warning
	{Citation `\@citeb' on page \thepage \space undefined}}
	{\csname b@\@citeb\endcsname}}}{#1}}

\newif\if@cghi
\def\cite{\@cghitrue\@ifnextchar [{\@tempswatrue
	\@citex}{\@tempswafalse\@citex[]}}
\def\citelow{\@cghifalse\@ifnextchar [{\@tempswatrue
	\@citex}{\@tempswafalse\@citex[]}}
\def\@cite#1#2{{$\null^{#1}$\if@tempswa\typeout
	{IJCGA warning: optional citation argument 
	ignored: `#2'} \fi}}

\def\@refcitex[#1]#2{\if@filesw\immediate\write\@auxout
	{\string\citation{#2}}\fi
\def\@citea{}\@refcite{\@for\@citeb:=#2\do
	{\@citea\def\@citea{, }\@ifundefined
	{b@\@citeb}{{\bf ?}\@warning
	{Citation `\@citeb' on page \thepage \space undefined}}
	\hbox{\csname b@\@citeb\endcsname}}}{#1}}

\def\@refcite#1#2{{#1\if@tempswa\typeout
        {IJCGA warning: optional citation argument
	ignored: `#2'} \fi}}

\def\refcite{\@ifnextchar[{\@tempswatrue
	\@refcitex}{\@tempswafalse\@refcitex[]}}

\def\pmb#1{\setbox0=\hbox{#1}
	\kern-.025em\copy0\kern-\wd0
	\kern.05em\copy0\kern-\wd0
	\kern-.025em\raise.0433em\box0}

\def\fnt#1#2{\footnotetext{\kern-.3em
	{$^{\mbox{\scriptsize #1}}$}{#2}}}

\font\tenrm=cmr10
\font\tenit=cmti10 
\font\tenbf=cmbx10
\font\bfit=cmbxti10 at 10pt
\font\ninerm=cmr9

\font\eightrm=cmr8

\def\qed{\hbox{${\vcenter{\vbox{			
   \hrule height 0.4pt\hbox{\vrule width 0.4pt height 6pt
   \kern5pt\vrule width 0.4pt}\hrule height 0.4pt}}}$}}

\begin{document}



\normalsize\textlineskip
\thispagestyle{empty}
\setcounter{page}{1}


\vspace*{0.88truein}

\centerline{\bf HAWKING-LIKE EFFECTS AND UNRUH-LIKE EFFECTS: TOWARD
EXPERIMENTS ?}
\vspace*{0.035truein}
\vspace*{0.37truein}
\centerline{\footnotesize HARET C. ROSU}
\vspace*{0.015truein}
\centerline{\footnotesize\it Instituto de F\'{\i}sica,
Universidad de Guanajuato, Apdo Postal E-143, Le\'on, Gto, Mexico}
\baselineskip=10pt
\vspace*{10pt}
\vspace*{0.225truein}

\vspace*{0.21truein}
\abstracts{
The Hawking effect and the Unruh effect are two of the most important
predictions in the theoretical physics of the last quarter of the
20th century. In parallel to the theoretical investigations there is great
interest in the possibility of revealing effects of this type in some sort
of experiments.
I present a general discussion of the proposals to measure the Hawking 
and Unruh effects and/or their `analogues' in the laboratory,
and I make brief comments on each of them. The reader may also find the
various physical pictures
corresponding to the two effects which were applied to more common phenomena,
and vice versa.
}{}{}


\textlineskip                  
\vspace*{12pt}                 

\vspace*{1pt}\textlineskip	
\vspace*{-0.5pt}
\noindent


\noindent




\noindent

\vskip 0.5cm



















\section{ Introduction}        

The most famous scientific formula seems to be $E = mc^{2}$,
which was settled by Einstein. However,
the following more recent formula is at least of the same rank

$$
T =\frac{\hbar}{2\pi c k}\cdot a
\eqno(1.1)
$$
including three fundamental physical
constants, and the rather geometrical constant $\pi$.
T is a quantum field temperature parameter identified with the
thermodynamic temperature and
$ a$ is the acceleration parameter in the physical system under
investigation. It was Hawking,\cite{haw} in 1974, who first obtained an
equivalent of Eq. (1.1) for Schwarzschild
black holes \footnote{For a concise look at black hole physics before the
Hawking effect, see Ref.[\refcite{ru}].},
being followed in 1975 by Davies,\cite{dav} who
provided a discussion of the quite
similar scalar particle production phenomena in Rindler
metric using a mirror model, and as a matter of fact, one can find Eq.(1.1)
clearly derived in 1976 in a seminal paper of Unruh,\cite{un}
who applied the so-called quantum/particle detector method.
The acceleration parameter is
the surface gravity in the black hole case ($\kappa =c^{4}/4GM$), and the
proper acceleration of a quantum system (most often the electron) in the
latter case. The radiated spectrum in these two cases may be considered
 as exactly of black body type (up to some distorsion due to the transmission
through the surrounding potential barrier for black holes), almost
 forcing one to accept a truly thermal/thermodynamic interpretation
 as very natural. However, other viewpoints may be favoured, e.g.,
 coherent effects,\cite{free,hil,lee,lee1,lon}
 or even better, squeezing effects of a quantum field
 vacuum,\cite{gri};
  Casimir-like effects,\cite{nug}; ambiguity in defining positive and
 negative modes,\cite{pah}; instanton effects,\cite{chr}.
There are standard methods to get these fundamental thermal-like field
effects, e.g., Euclidean Green's functions,\cite{gib,gibb}
Bogolubov mixing coefficients,\cite{haw,la}
construction of individual wave packets,\cite{wald} renormalized
energy-momentum tensor,\cite{dfu}
 instanton techniques,\cite{chr} analytic mappings,\cite{san}
 thermofield dynamics,\cite{is} and even more classical
 derivations,\cite{boy} (these latter ones may be considered to be at
 variance with the relationship between the Hawking effect and the Weyl trace
 anomaly.\cite{cf}) These effects
 are very general, since they are a direct consequence of the
 Fulling nonuniqueness of canonical quantization in curved
 spacetimes.\cite{ful} They must appear in the framework of any
 quantum/stochastic field theory, perhaps with some specific non-trivial
 features.
Various aspects of these effects have been revealed in the vast
literature that has been accumulated over the years; the interested reader
is directed
to some well-known review papers.\cite{dew,ish,sci,tak,full,gin}
 However, we have to point out
that we still lack a dedicated book although these effects
are discussed with various degree of detail in black hole and/or
quantum field theory
in curved space-time books, as well as in general relativity books
 (e.g., Frolov and Novikov,\cite{fron}
Caltech- membrane book,\cite{cal} Birrell and Davies,\cite{bid}
Wald.\cite{wa1})
 Probably we should wait first for a dedicated conference/symposium.
Perhaps, it is interesting to recall the concept of {\em thermal radiation
from nothing} elaborated by Kandrup,\cite{ka1} in the cosmological
context, which one may also take into account for the case of Hawking
radiation and Unruh radiation, if {\em nothing} means zero point energy.
In fact, Hawking and Unruh effects may be considered (and have been
considered) as {\em intelectual surprises} in the sense given by
Peierls,\cite{pei}
i.e., they could have been predicted much earlier (for how much earlier
there are different opinions!).
 
The main goal of this survey paper is to present
the experimental settings that were proposed so far to detect
such thermal-like effects, attaching short comments to each of
them. In cgs system Eq. (1.1) turns into:
$$
T=4\times 10^{-23}\cdot a
\eqno(1.2)
$$
and therefore we need accelerations greater than $10^{20}g_{\oplus}$
 ($g_{\oplus}$ is the mean Earth surface gravity)
 in order
to have a `heat-bath' quantum vacuum at the level of only one Kelvin.
We are facing extremely small thermal-like quantum noises
requiring on the scale of terrestrial laboratories extremely strong
non-adiabatic perturbations to be applied.
That is, such perturbations are not built up at a constant rate from
zero to their final value over a time interval which is long as compared
to inverse quantum frequencies, and therefore one is changing the
occupation probabilities, i.e., directly the populations of the quantum states.
Non-adiabatic transitions are a necessary prerequisite of all sorts of
heat, including those to be discussed next.
Generally, the Hawking effect is considered to show up in the realm of
astrophysics, whereas the Unruh effect is more appropriate to
an extremely non-perturbative and non-linear electrodynamic regime.\cite{mcd}
Indeed, the radiated Unruh power is $4.1\times10^{-118}a^4$ as compared
with the Larmor power which is $5.7\times10^{-51}a^2$ in cgs units
and one could see that the two radiations are comparable for $a=
3\times10^{30}g_{\oplus}$.
 Such accelerations are produced by electric and/or
magnetic fields that are one order of magnitude beyond the critical
electrodynamical field
$F=m^{2}c^{3}/e\hbar\approx 1.32\times 10^{16}V/cm\approx 4.41\times10^{13}G$
at which the spontaneous electron-positron
pair production in vacuum starts on. Thus Unruh effect will
be competed by non-linear electrodynamic effects
(first of all nonlinear Thomson scattering.\cite{mcd}) On the other
hand, Nikishov and Ritus,\cite{nik} on account of processes induced by
charged particles in an electric field,
provided arguments against the Unruh heat-bath
concept. Their point is that on the length scale of quantum pair-production
processes, one cannot encounter a constant acceleration field,
so that in general the occuring pair production processes are not
thermal/Unruh-like. In their paper, they worked out probability exponents for
various cases of pair production in a constant electromagnetic field and
showed that as a rule these are not `thermodynamic' ones, i.e., linear in the
excitation energy. For instance, they considered the probability of the process
$q_{1}\rightarrow q_{2}+q_{3}$ in the case of weakly differing accelerations
of the products, and obtained a thermodynamic exponential with an effective
excitation temperature which is twice the Unruh temperature. This is why
they called such a case an ideal detector of uniformly accelerated motion.
It is also quite well-known the opinion of these authors that Rindler states
cannot be produced in Minkowski space without sources of infinite power on
the event horizons. Another argument against the physical realization of
coordinate systems such as the
Rindler one was provided by Padmanabhan,\cite{pad} who, by analogy with
QED, concluded that any physically realizable coordinate system can differ
from the Minkowski coordinates only in a finite region of spacetime.

One should also recall the debated point of extrapolating
Lorentz invariance to extremely high energies that was emphasized by Jacobson
in the black hole case.\cite{jac} In fact, it is well known
that the black body
spectrum is the only distribution that is Lorentz and even conformal
invariant. Indeed, the usual Bose distribution is just the form in the
rest frame of the Lorentz scalar $1/[\exp(\beta pv)-1]$, where $v$ is
the four velocity of the thermal bath.

With the purpose of relating Hawking effect and Unruh effect
to Everyday
Physics/Earth laboratories, we shall use the most
simple/intuitive vocabulary at our disposal throughout the paper.

To close the Introduction, a comment about the title.
By Hawking-like and Unruh-like effects we mean effects of Hawking and Unruh
type, i.e., quite
{\em similar} effects which need only {\em minor} modifications
(in my opinion, the latter condition is not well satisfied in some cases
I'll be reviewing next).
In order to
test scientific ideas that are applied in domains much beyond our present
technologies, we have to find equivalents to those ideas that might be
tested in the laboratory. One can encounter quite a few
papers
written with this purpose in the literature.\cite{bow} In this regard, the
list of `experimental' proposals to follow is a clear illustration
of such an idea. At the same time, the
inverse action is also at one's disposal, that is, applying frameworks of
ordinary effects to Black Hole Physics.\cite{gid}

The organization of the paper is as follows: the next section contains
a short presentation of the original derivations of the effects;
all sections thereafter deal with the
analogs and the model experiments suggested for Hawking and Unruh
effects. Although not exhaustive, a quite broad material is brought together
in order to imprint on the reader what might be a useful global view on many
topics.

\vspace*{0.2cm}

\section{Hawking's and Unruh's Paradigms}     

This section is included in the review for self-consistency reasons, otherwise
it follows closely published text.

The remarkable results of Hawking and Unruh belong, together with the
Penrose effect,\cite{pen} and the
superradiance,\cite{sta} to the `epoch of effects'
that occured in black hole physics in early 1970s.
At the present time, these effects are standard theoretical paradigms
in quantum field theory, astrophysics, and cosmology. In this section we
remind the basic ideas used by the two authors in their seminal papers on
the topic.

During 1974-1976, Hawking dealt with the Klein-Gordon equation for a massless
scalar
field in a Schwarzschild metric and he used naturally an in-out formalism for
radial null rays that define a one-to-one mapping between past null infinity
and future null infinity as
required by an asymptotic observer. Since the usual coordinates of the
background become singular
at the horizon, one should consider there the Kruskal-Szekeres regular set,
which
is related to the Schwarzschild coordinates by transformations reminding
the Langer transformation in ordinary semiclassical physics. Hawking
propagated the scalar normal modes by the method of geometrical optics and
used the
method of Bogolubov coefficients (at that time already well established
in quantum
cosmology/cosmological particle production, for a review see
Ref.[\refcite{par}])
to obtain his famous conclusion that
black holes are thermal objects. An incoming null ray $v$= const, originating
on $\cal I ^{-}$ propagates through the gravitational background  to become
an outgoing null ray $u$= const, arriving on $\cal I ^{+}$ at a value
$u=F(v)$. Similarly, and this was the procedure of Hawking, one can trace a
null ray from a constant $u$ on $\cal I ^{+}$ to a $v = G(u)$ on $\cal I ^{-}$,
where the function $G$ is the inverse of $F$. Getting one of these
functions is the clue towards all the physical results.
By this geometrical ray-tracing (or constant phase tracing),
Hawking obtained in the Schwarzschild case (Newton, Planck, Maxwell, Boltzmann
constants all set equal to unity)
$$
G(u)=-C\exp [(4M)^{-1}u] + v_{0}
\eqno(2.1)
$$
where $C$ and $v_{0}$ are constants. The latter constant
denotes the ingoing ray
that reaches the horizon at the moment of its formation. For quantization,
one needs complete sets of mode functions, that is ingoing and outgoing wave
packets constructed from massless spherical waves.\cite{haw} The wave packets
are solutions of the scalar equation in the Schwarzschild background geometry
if the gravitational backscattering is neglected. Even so, the exponential
increase with advanced time of the mean frequency of the wave packets at
$\cal I ^{-}$ is a well-known disturbing feature in the intermediate stages
of the calculation.\cite{wa1} The quantization, carried on wave packet mode
functions on the two asymptotic infinities, brings in wave packet creation and
annihilation operators and the Bogolubov mixing coefficients can be calculated
using the ray tracing formula to  change conveniently the variables. For a
recent presentation of the formalism with interesting discussions of the
localization issue see.\cite{am} The main result is the
following relation between the two Bogolubov coefficients
$$|\alpha _{\omega \omega ^{'}}|^{2}=e^{2\pi \omega/\kappa}
|\beta _{\omega \omega ^{'}}|^{2}
\eqno(2.2)
$$
which is exactly the relation required at the level of Bogolubov coefficients
to obtain the emission of a blackbody spectrum.

By a straightforward application of the Hawking formalism to the Rindler
wedge equipped with a reflecting mirror placed to the right of the origin,
Davies,\cite{dav} obtained the result that an observer moving with
uniform acceleration
$a$ sees the fixed surface of the mirror radiating a thermal spectrum with a
temperature $a/2\pi$, that may be considered as a close variant of the Unruh
effect.

An essential mathematical result that one needs in order to discuss the
Unruh effect refers to writing the Minkowski vacuum state as an entangled
Rindler vacuum state
$$
|0\rangle _{M}=
C\Big[\prod _{i}\exp(e^{-2\pi\omega _{M}}r^{\dagger}_{i}l^{\dagger}_{i})\Big]
|0\rangle _{R}
\eqno(2.3)
$$
where $C$ is a normalization constant, the indicial set $(i)$ is
$(i)=(\omega_{R}, k)$, the creation operators $r^{\dagger}$ and
$l^{\dagger}$ live only in the
right and left Rindler wedge respectively, and the other subscripts are obvious.
Gerlach has written interesting papers on this
vacuum `superfluidity'.\cite{ger}

In the third section of his ``Notes on black-hole evaporation"
Unruh,\cite{un} presented a clear analysis of the behavior of
particle/quantum detectors
under acceleration in flat spacetime with two remarkable findings

(i) {\em A particle detector will react to states which have positive
frequency with respect to the detector proper time, not with respect to any
universal time}.

(ii) {\em The process of detection of a field quanta by a detector, defined as
the exciting of the detector by the field, may correspond to either the
absorption or the emission of a field quanta when the detector is an
accelerated one}.

These fundamental conclusions are reached by investigating two types of
detectors. The first one, a box containing a Schr\"{o}dinger particle in its
ground state,
which is said to have detected a quanta of a massless scalar field if the
detector is found in a state other than its ground state at some late time.
 The second one is a relativistic model describing the
	interaction of two complex scalar fields, one of which is
	the detector field $\Psi$ of mass $M_{d}$ coupled via a real
	scalar field $\Phi$ to an `excited' state representing the
	second scalar field $\phi$ of mass $M>M_{d}$. The detector field
	$\Psi$ is said to have detected a $\Phi$ quantum if it changes
	into the second, excited field $\phi$ at some late time.

The box detector is held fixed on a line of constant acceleration $a$ in
Rindler space (or else, the box moves with constant linear acceleration in
Minkovski space).
In this case the Schr\"{o}dinger equation for the particle in
the box has an additional potential  $m\zeta a$, where
$\zeta = (2a)^{-1}[4a^{2}\rho-1]$ is the proper
length coordinate in the Rindler polar coordinate $\rho$,
assumed small within the box. Unruh performed a calculation
of the lowest-order transition rate per unit proper time for an interaction
of the form $\epsilon \Phi \psi $, where $\psi$ belongs to the set
$\psi _{j} =\exp (-iE_{j}\tau /2a)$ of
detector states. In addition, expanding $\Phi$ in Rindler modes yields a result
containing two factors. One may be interpreted to represent the
destruction of one of the Rindler particles in the product of Rindler states
which exist in the Minkowski
vacuum (see Eq. (2.3)) by the detection process. The other factor
represents the
`sensitivity' (or efficiency) of the detector to a Rindler mode.
Unruh obtained the final
result (the vacuum heat bath) by passing to Minkovski creation and annihilation
operators, and by evaluating the `sensitivity' in a WKB approximation.
The result essentially comes out from the fact that the detector measures
frequencies with respect to its proper time which is not the same for all
geodesic detectors in accelerated reference systems.

Before ending this section, perhaps it is worthwhile to mention the
association of thermal effects with
the `above barrier' reflection coefficient that has been noticed only by few
authors.\cite{sa,gri} The point is that the semiclassical
`above barrier' reflection coefficient is exponentially small and therefore
allows the emergence of `thermal' backscatterings. I have commented on
the importance of this standpoint elsewhere.\cite{r93}
It allows the interesting development of considering
new thermal regimes for black holes within the WKB approach with two
turning points. In solid state physics these WKB thermal effects are,
{\em e.g.}, field emission and thermionic emission of electrons from solid
surfaces.

\vspace*{0.2cm}

\section{ Hawking Effect in Astrophysics}   

Hawking radiation is insignificant for stellar mass black holes and
only primordial black holes (PBHs), i.e., those having a mass
smaller than $M_{c}=10^{15}g$ (this is the mass of a common Earth mountain)
could have a detectable Hawking luminosity. The point is that
as early as 1976 Hawking and Page,\cite{haw2} concluded that mountain-mass
black holes (they would have to be hadron-sized objects)
formed in some way in the very early Universe
(phase transitions, bubble collisions, string collapses, Zel'dovich-
 Harrison density perturbations) are at the present epoch in their
 final evaporation stages (denoted as Hawking explosions).
  The temperature of the PBHs
 posessing the critical mountain-mass is around 14 MeV, and they have
 been emitting on a time scale comparable with the lifetime of the
 Universe at the peak black body photon radiation located at 14 MeV.
 Consequently, the most
 simple experiment is to measure the photon flux
 in the tens of MeV range by means of
 satellite-borne detectors. This has been done already in 1977-1978
 but the measured $\gamma$ flux was seen to fall with energy as
 $E^{-2.5}$ without any evidence for a photon excess in the vicinity
 of 14 MeV.\cite{fst} The negative result was turned into a well-known
limit on the number of PBHs per logarithmic mass interval at the critical
mass, the so-called Hawking-Page [HP] bound of 1976
 $$
 N=\frac{dn}{d(\ln\; M)}|_{M=M_{c}}<10^{5}\;(10^{11})\;pc^{-3}
 \eqno(3.1)
 $$
 and the HP explosion rate
 $$
 \frac{dn}{dt}=\frac{3\alpha (M_{c})}{M_{c}^{3}}\;N=2.2\cdot 10^{-10}\;N\;
 pc^{-3}\;yr^{-1}~.
 \eqno(3.2)
 $$
 The latter, however, is strongly
 dependent on the cosmological parameters (spacial curvature and
 Hubble $h_{0}$ parameter), the astrophysical
 premises (e.g., that PBHs have clustered- second figure in Eq. (3.1)- or not
 into galaxies- first figure in Eq. (3.1)), and the particle
 physics parameter $\alpha (M_{c})$ which is a measure of the
 degrees of freedom coming from particle physics.

In 1989, Halzen and Zas,\cite{hal}
 reanalyzed the MeV limit on the number of critical PBHs by
 taking into account the particle degrees of freedom of the
 standard model of quarks and leptons. They obtained an increase
 of one order of magnitude in their density Eq. (3.1), and of two orders of
 magnitude in their explosion rate Eq. (3.2). On an intuitive base, the revised HP
 limit says that there cannot be more than about 2000 explosions per
 $pc^{3}$ per year assuming galactic clustering of the critical PBHs in our
 galactic neighborhood.
 At higher energies, the observed spectrum from PBHs
 is a convolution of the fundamental emission spectrum with the quark
 fragmentation functions, resulting in a power law at energies above a
 few TeV in the last seconds before explosion. It is worth noting however
 that multiparticle production at accelerators revealed that gluonic
 branching processes may well be dominant over quark branching.\cite{bm}. In
 order to obtain reliable theoretical results on the
 extremely high-energy spectrum emitted by PBHs, one needs insights into
 the general mathematical theory of branching processes as applied to
 multiparticle production.\cite{mp}

Apparently, there exist real possibilities for detecting
 Hawking bursts in the TeV and PeV range to sky depths not excluded by the
 HP bound by means of the new generation of air shower arrays (e.g., CYGNUS,
 CASA) and
 Cherenkov telescopes (e.g., the Whipple Telescope on Mount Hopkins, Arizona).
For further details we refer the reader to some
literature.\cite{hal3,macg,macg1,macg2,macg3,gril,ram}

The last data taken with the CYGNUS detector between 1989 September
and 1993 January in search for one-second bursts of ultrahigh energy gamma
rays
from arbitrary located point sources in the northen sky finds no evidence
for such bursts. Moreover, it sets the most restrictive upper limit at the
moment of
$8.5\times 10^{5} pc^{-3}yr^{-1}$ at the confidence level of 99\% for the
rate-density of evaporating PBHs, assuming them uniformly distributed in
the Sun neighborhood.\cite{ale}
The CYGNUS detector is located in Los Alamos, New Mexico,
and consists of 108 scintillation counters of 1 $m^{2}$ each, deployed over
22000 $m^{2}$. The mean primary energy for gamma-ray-initiated events is
50 TeV. The information provided by the CYGNUS array is gathered from a
larger volume of space, being complementary to
that of atmospheric Cherenkov telescopes which are able to probe greater
distances.

In the same astrophysical context, suppose that one day we will be almost sure
that a certain object
or group of objects are black holes, perhaps surrounded by some material
shells. Of course, we shall have
some sort of power spectra from them and we would like to
determine the horizon area temperature distribution. Taking the black body
origin as granted (it can be
argued that the overhorizon correlations are precisely such that the
spectrum comes out as of a black body),
we will face the inverse black/grey body problem for a quantum (horizon)
lightlike surface.\cite{tho} This problem is not at all trivial even for
classical surfaces. Some hints may be found in some of my works, where I
considered it as an {\em inverse Moebius problem},\cite{moe} on the lines
of the developments due to N.-x. Chen,\cite{bb}. Also, the
coherence characteristics of black hole sources are basically unknown at the
present time, although some conclusions may be drawn from the squeezing
picture of the Hawking radiation.

To conclude this section, we remind Chapline's discussion on the connections
of PBHs and hadron physics.\cite{cha} A number of authors
have studied low mass (mini) black holes in the particle physics
perspective.\cite{vis} Also, Turner's old question whether
PBHs might be the source of the cosmic ray antiprotons,\cite{tur} has been
reanalysed by MacGibbon \cite{macg3}. In the same cosmic -ray context,
Greenberg and Burns,\cite{gb}
commented on the ionization tracks and ranges of small black holes, however
without taking into account the Hawking radiation. This might work in the case
of some kind of black hole relics/remnants. As a matter of fact, this
last paragraph may be thought of as a connection with Section 13 below.
 
\vspace*{0.2cm}

\section{ Hydrodynamical Hawking Effect}      

 In a Physical Review Letter of 1981, Unruh showed that
 a thermal spectrum of sound waves should be given out from the
 sonic horizon/Mach shock wave in transsonic fluid flow.\cite{un1} Starting
 with the equations of motion of an irrotational fluid (i.e., Navier-Stokes
 and the continuity equation) and
 linearizing them, the perturbations of the flow can be described
 by a massless scalar field in a metric determined by the
 background fluid velocity field. This metric looks like a Schwarzschild
 metric when a spherically symmetric,
 stationary convergent flow exceeding smoothly
 the speed of sound at some radius (the horizon radius) is considered.
 That is, in the radial outward direction the velocity of sound is
 $$
 v_{r}=-c+\alpha (r-R) +{\cal O}((r-R)^{2})
 \eqno(4.1)
 $$
 and the radial part of the background fluid metric is
 $$
ds^{2}=\frac{\rho _{0}(R)}{c}\Big[2c\alpha(r-R)d\tau ^{2} -
\frac{dr^{2}}{2\alpha(r-R)}\Big]
\eqno(4.2)
$$
where $\alpha =\frac{\partial v}{\partial r}$ is the radial velocity gradient.
 One may recognize this metric due to the velocity/stream potential
 as similar to the metric near a Schwarzschild horizon.
After quantising the sonic comoving field, Unruh finds near the sonic horizon
the following time dependence of the phonon modes
$$\phi _{\omega}\approx (t-t_{0}+const)^{i\omega /\alpha}   \eqno(4.3)$$
which is similar to the $(t-t_{0})^{i\kappa \omega}$ dependence near the
Schwarzschild horizon with respect to a freely falling observer.
The thermal flux of phonons would be at a temperature
$$
T=\frac{\hbar}{2\pi k}\cdot\frac{\partial v}{\partial r}\simeq
 10^{-2}K\cdot\frac{\partial v}{\partial r}
\eqno(4.4)
$$

 Since the transsonic transition is usually accompanied by
 turbulent instabilities, one would expect  the sonic thermal-
 like spectrum at the spherical shock to be much under any experimental
 detection. Indeed, in order to have a Planck spectrum peaked at only
1 K, the gradient of the velocity at the shock has to reach $100$ m/s per \AA.
This estimate is very disappointing. It is almost sure that a simple
atomic fluid cannot support such huge gradients. However, the situation
may change in the case of superfluids, as first argued by Comer.\cite{Com}
Already in the summer of 1991, Volovik,\cite{svolo} wrote a paper with
Schopohl on the analogy between Schwinger pair production and superfluidity
($^3$He-B) and he
is actively pursuing the analogy project between quantized vortices
and black holes \cite{volo}. For other important
discussions of quasiparticle pair creation in unstable superflow,
the reader is directed to Elser's papers.\cite{els}

Jacobson,\cite{jac} discussed the fluid flow analogy
in the context of the question ``would a black hole radiate
if there were a Planck scale cutoff in the rest frame of the hole?''.
Trying to give an answer, Jacobson developed an
interesting {\em superfluid black hole} model
which certainly has to be further elaborated.\cite{rld} Indeed, one may
work out hybridization mechanisms between surface and bulk modes
({\em ripplons}
and {\em rotons}) in what may be an attractive physical picture for
subtle problems in black hole physics.

It is worthwile to recall also
the {\em Planck aether substratum/ vortex sponge}
of Winterberg,\cite{win} which may be useful
despite the ancient (19th century) quaint picture. A great deal of superfluid
and vortex turbulence literature may also be looked upon in the above
perspective.


The Mach horizons deserve further studies from the point of view of
their thermal-like effects, because together with Cherenkov horizons, are
the closest material structures to the black hole lightlike horizons
one can think about.

Furthermore, since collapse may be reduced to appropriate boundary conditions
on the past horizon (see Unruh's ``Notes'' of 1976), more should be known on
outgoing boundary conditions
for dispersive waves in hydrodynamics. A good paper on these lines belongs
to Israeli and Orzsag.\cite{io}

In addition, Hayward has recently discussed an outgoing
spherically symmetric light-cone evolving according to the
Einstein equations.\cite{sah}
Hamiltonian formulations for relativistic superfluids should be taken into
account as powerful formalisms for investigating phenomena of Hawking and
Unruh type.\cite{cl}

\vspace*{0.2cm}

\section{Unruh Effect in Storage Rings}

J.S. Bell (`the quantum engineer') and his co-workers,
J.M. Leinaas and R.J. Hughes,
have imagined another experimental scheme connected to the
Unruh effect. During 1983-1987 they published a
number of papers on the idea that the depolarising effects in
electron storage rings could be interpreted in terms of Unruh
effect.\cite{bell,bell1,bel,mcd1} 
However, the incomplete radiative polarization of the electrons
in storage rings has been first predicted in early sixties in
the framework of QED. Besides, it is known that
the circular vacuum noise does not have the same universal thermal
character as the linear Unruh noise.\cite{tak} This appears as a
`drawback' of the `storage ring electron thermometry', not to
mention the very intricate spin physics. Keeping in mind these facts,
we go on with further comments, following the very clear discussion of
Leinaas.\cite{lei}

The circular acceleration in the LEP machine is $a_{LEP}=3\times
10^{22}g_{\oplus}$
corresponding to the Unruh temperature $T_{U}=1200 K$. It is a
simple matter to show that an ensemble of electrons in a uniform
magnetic field at a nonzero temperature will have a polarization
expessed through the following hyperbolic tangent
$ P_{U}=\tanh({\frac{\pi G}{2\beta}})$.
For the classical value of the gyromagnetic factor ($G=2$) and for
highly relativistic electrons ($\beta=1$),
$P_{U}=\tanh{\pi}=0.996$, beyond the limiting polarization of
Sokolov and Ternov,\cite{sok} $P_{lim}=\frac{8\sqrt{3}}{15}=0.924$.

On the other hand, the function $ P_{U}(G)=\frac{1-e^{-\pi G}}{1+e^{-\pi G}}$
is very similar, when plotted, to the
function $ P_{DK}(G)$, which is a combination of exponential and polynomial
terms in the anomalous part of the gyromagnetic factor of the electron,
and it was obtained through QED calculations by Derbenev
and Kondratenko.\cite{der}
 The difference is merely a shift of the latter
along the positive G-axis with about 1.2 units. As shown by Bell
and Leinaas, when the Thomas precession of the electron is included
in the spin Hamiltonian a shift of 2 units is obtained for $P_{U}(G)$.
This suggests a more careful treatment of spin effects arising when
one is going from the lab system to the circulating coordinate frame.
A new spin Hamiltonian was introduced by Bell and Leinaas with a more
complicated structure of the field vector in the scalar product
with the Pauli matrices. This complicated structure takes into
account the classical external fields, the quantum radiation field
and the fluctuations around the classical path. Within this more
complete treatment, Bell and Leinaas were able to get, to linear order
in the quantum fluctuations, a Thomas-like term and a third resonant
 term directly related to the vertical fluctuations in the electron
 orbit, which are responsible for the spin transitions.
 The resonance factor, denoted $f(G)$, induces an interesting variation of the
beam polarization close to the resonance. As $ \gamma$ passes through it
from below, the polarization first falls from $92\%$ to $-17\%$, and then it
increases
again to $ 99\%$ before stabilizing to $92\%$ . This is the only clear
difference from the standard QED. Such resonances induced by the
vertical fluctuations of the orbit have been considered
 before in the Russian literature but focusing strictly on their
 depolarizing effect. Their nature is related to the fact that
 the Fourier spectrum of the energy jumps associated with the
 quantum emission processes contains harmonics giving the usual
 resonance condition. As emphasized by Bell and Leinaas, a more direct
 experimental demonstration of the circular Unruh noise would be
 the measurement of the vertical fluctuations. However, this will clearly
be a very difficult task since such fluctuations are among the
  smallest orbit perturbations. At the same time, the measurement
  of the polarization variation close to the narrow resonance, in
  particular the detection of polarizations exceeding the limiting
  one, will make us more confident in the claims of Bell and Leinaas.
  It is worth mentioning that the rapid passage through the resonance
  does not change the polarization, while a slow passage reverses
  it but does not change the degree of polarization. Therefore only
  an intermediate rate with respect to a quantum emission time scale
  of passing through the resonance will be appropriate.

Barber and Mane,\cite{bar1} have shown that the DK and BL
formalisms for the equilibrium degree of radiative electron
 polarization are not so different as they might look, and they also obtained
 an even more general formula for $P_{eq}$ than DK and BL ones. On the
 base of their formula they estimated as negligible the BL increase near
  the resonance.

The basic experimental data on spin depolarizing effects remain as yet
those measured at
SPEAR at energies around 3.6 GeV in 1983. Away from the resonant
$\gamma's$ the
maximum polarization of Sokolov and Ternov was confirmed.\cite{john}

We are going to address now some spin physics effects in external fields of
critical strength.
As was stated in the Introduction, Unruh effect may show up in such
fields as a kind of thermal background for some
 nonlinear phenomena with thresholds in that energy region.
 In the spin physics context the detailed structure of the
 electron mass operator M has to be known for such fields. We refer
 the reader to the paper of Bayer {\em et al},\cite{bai}
where one can find expressions for the real part of M (related to the
anomalous magnetic moment of the electron) and the imaginary part (related to
 the probability of emission).

Ternov,\cite{ter} provided
 a quantum generalization of the BMT evolution
equation including the effects of Zitterbewegung and of the gradients
of the magnetic field, expected to become important in the critical regime.

One should mention the quasiclassical trajectory
coherent states introduced by Bagrov and Maslov.\cite{bag} These
states have been used in obtaining another generalization of the
BMT equation for the electron spin in the case of an arbitrary external
torsion field.\cite{bag1}

A recent paper of Cai, Lloyd and Papini,\cite{reg} claims that the
Mashhoon
effect due to the spin-rotation coupling is stronger than the circular
Unruh effect (spin-acceleration coupling) at all accelerator energies
in the case of a perfect circular storage ring. However, the comparison
is not at all a straightforward one.

Bautista,\cite{bau} solved the
Dirac equation in Rindler coordinates with a constant magnetic field
in the direction of acceleration and showed that the Bogolubov coefficients
of this problem do not mix up the spin components. Thus there is no spin
polarization due to the acceleration in this case.

In our opinion, the real importance of considering Unruh effect at storage
rings is related to clarifying radiometric features of the synchrotron
radiation.\cite{ror} There is a strong need to establish radiometric standards
in spectral ranges much beyond those of the cavity/blackbody standards,
and synchrotron radiation has already been considered experimentally
from this point of view.\cite{ku} Quantum field thermality is
intrinsically connected to the KMS condition. This is a well-known
skew periodicity in imaginary time of Green's functions expressing the
detailed
balance criterion in field theory. However, the task is to work out in more
definite terms the radiometric message of
the KMS quantum/stochastic processes.\cite{kms}

\vspace*{0.2cm}

\section{ Unruh Effect and Geonium Physics}    

The very successful Geonium physics could help detecting the circular
thermal-like vacuum noise. The proposal belongs to J. Rogers,\cite{rog}
and apparently it is one of the most attractive.
The idea of Rogers is to place
a small superconducting Penning trap in a microwave cavity. A single
electron is constrained to move in a cyclotron orbit around the trap
axis by a uniform magnetic field (Rogers figure is B = 150 kGs).
The circular proper acceleration is $a= 6\times 10^{21}g_{\oplus}$
corresponding to T = 2.4 K. The velocity of the electron is maintained fixed
($\beta= 0.6$) by means of a circularly polarized wave at the electron
cyclotron frequency, compensating also for the irradiated power.
The static quadrupole electric field of the trap creates a quadratic
potential well along the trap axis in which the electron oscillates.
The axial frequency is 10.5 GHz (more than 150 times the typical
experimental situation,\cite{bg}) for the device scale chosen by
Rogers. This
is the measured frequency since it is known,\cite{bg} that the best way
of observing the electron motion from the outside world (Feynman's ``rest of
the Universe'') is through the measurement of the current due to the induced
charge on the cap electrodes of the trap, as a consequence of the axial
motion of the electron along the symmetry axis.
At 10.5 GHz the difference in energy densities between the circular
noise and the universal linear noise are negligible (see Fig. 2 in
Rogers' work). Actually, Rogers used the parametrization for the spectral
energy density of a massless scalar field as given by Kim, Soh
and Yee.\cite{ksy} Their calculation is based on the Wightman
two-point functions
(recall that in quantum optics this is equivalent to not assuming the
rotating wave approximation) and yields the following result:
$$
\frac{de}{d\omega}=\frac{\hbar}{\pi ^{2}c^{3}}\Big[\frac{\omega ^{3}}{2}+
\gamma \omega _{c}^{3}x^{2}\sum _{n=0}^{\infty}\frac{\beta ^{2n}}{2n+1}
\cdot\sum _{k=0}^{n}(-1)^{k}\frac{(n-k-x)^{2n+1}}{k!(2n-k)}\Theta [n-k-x]\Big]
\eqno(6.1)
$$
where $\gamma$ is the relativistic gamma factor,
$x=\omega/\gamma \omega _{c}$, $\omega _{c}=eB/\gamma mc$ is the
cyclotron frequency, and $\Theta$ is the Heaviside step function. 
The power spectral density at the axial frequency is only
${\partial P}/{\partial f} = 0.47\cdot 10^{-22}$ W/Hz, and may be assumed
to be almost the same as the electromagnetic spectral energy density. This
power is resonantly transfered to the $TM_{010}$ mode of the
microwave cavity and a most sensible cryogenic GaAs field-effect
transistor amplifier should be used to have an acceptible
signal-to-noise ratio of S/N = 0.3. According to Rogers, the signal
can be distinguished from the amplifier noise in about 12 ms.

In conclusion, very stringent conditions are required in the model
experiment of Rogers. Top electronics and cryogenic techniques
are involved as well as the most sophisticated geonium methods. Taking
into account the high degree of precision attained by geonium
techniques, one may think of Rogers' proposal as one of the most feasible.
The critique of this proposal is similar to that in storage rings,\cite{ros}
namely that the circular Unruh effect is not universal, depending also on the
electron velocity. Also, Levin, Peleg, and Peres,\cite{lpp} studied the
Unruh effect for a massless scalar field enclosed in a two-dimensional circular
cavity concluding that the effects of finite cavity size on the
frequencies of normal modes of the cavity (Casimir effect) ignored by Bell
{\it et al}, and by Rogers are in fact quite important.

A better experimental setting for detecting vacuum noises
by means of a trapped quantum detector (electron) may well be the cylindrical
Penning trap,
for which the trap itself is a  microwave cavity.\cite{tg} In this case
small slits incorporating choke flanges divide high-conductivity copper
cavity walls into the required electrode Penning configuration, including two
compensation electrodes. The driven axial resonance for this configuration
has already been observed with almost the same signal-to-noise ratio as
in hyperbolic Penning traps. By means of these cylindrical cavity traps,
a more direct coupling to the cavity modes may be achieved, especially in the
weak coupling regime, where the cyclotron oscillator and the cavity mode
cannot form normal modes, and therefore other nonlinear effects are not coming
into play. The cylindrical $TM_{010}$ mode is essentially a zero-order Bessel
function in the radial direction and has no modes along the z axis. The price
to pay in the case of the cylindrical trap is a loss of control on the
quality of the electrostatic quadrupole potential.

\vspace*{0.2cm}

\section{ Hawking Effect and Casimir Effect}     

There exist strong similarities between Hawking-like effects
(black hole physics in general)
and the Casimir effect. Indeed, the global structure of the
spacetime manifold is what really matters for Hawking-like
effects, and makes the general features of Hawking's result
to be met already in the moving mirror models.

The Hawking effect might be looked upon as a Casimir effect if
one argues as follows.
 The causal constraints generate peculiar surfaces
(horizons) that may be considered as some kind of boundaries.
Price,\cite{pri} and
Fabbri,\cite{fab} have shown long ago that the gravitational
field of a black hole creates an effective potential barrier
acting as a good conductor in the low frequency range and
blocking the high multipoles of the high frequency electromagnetic waves.
The barrier is very well
localized near $r = 1.5 R_{h}=3M$. For low frequency waves (
$\omega \leq \omega _{c}= (2/27)^{1/2}M^{-1}$) there are two real turning
points for all partial waves). According to Nugayev,\cite{nug}
who elaborated in more detail
on the analogy between the Hawking effect and the Casimir effect, all
the `thermal'
radiation is born in the spatial region between the two turning points.
Nugayev's goal was to
investigate the `particle production' by a black hole in terms of
temperature corrections to the Casimir effect, which
are due to the interaction of the radiation with the surface of the
potential barrier. He claimed that the two turning points are at different
temperatures ($T_{inner}> T_{outer}$) and therefore a (Casimir) energy flow
from the inner to the
outer turning spherical region should occur.
This flow of Casimir energy makes the
area of the horizon to shrink at precisely the rate consistent with the
energy flux at spacial future infinity. Nugayev also argues that the virtual
particles are turned into real ones by the small but infinite tail of the
potential barrier beyond the maximum which lies at $3M$.

Making use of the Casimir picture, Nugayev predicted in another work two
regimes of black hole evaporation, an anomalous skin effect
regime at low temperatures and a normal skin effect at higher
 temperatures.\cite{nug1}

However, one should keep in mind that
no analogy is complete. As was first shown by Ford,\cite{for}
 although the vacuum energy density in bounded space may
have thermal representations (see also Ref.[\refcite{co}]),
 the spectrum of the Casimir effect
is not at all thermal. This may be seen when one is revealing
the contribution of each frequency interval to the Casimir energy by
means of weighting functions, as Ford has proven.

Let us also mention the connection found by Burinskii,\cite{bur} who
developed a model for the material of the Kerr-Newman metric source based on
the usually neglected {\em volume} Casimir energy.

\vspace*{0.2cm}

\section{ Unruh Effect and Nonadiabatic Casimir Effect}  

An experimental equivalent of a fast moving mirror might be a
plasma front created when a gas is suddenly photoionized. This is
the proposal of Yablonovitch.\cite{yab} The argument is that
 the phase shift of the zero-point electromagnetic field
transmitted through a plasma window whose index of refraction
is falling with time (from 1 to 0) is the same as when reflected
from an accelerating mirror. Consider the case of hyperbolic motion.
Since the velocity is
$$
v= c\tanh(a\tau/c)
\eqno(8.1)
$$
where $\tau$ is the observer's proper time, the Doppler shift frequency
will be
$$
\omega_{D} = \omega_{0}\sqrt{\frac{1 - v/c}{1 +v/c}} =
\omega_{0}\exp({-a\tau/c})
\eqno(8.2)
$$
and consequently a plane wave of frequency $\omega_{0}$ turns into
a wave with a time -dependent frequency. Such waves are called chirped
waves in nonlinear optics and acoustics. Eq. (8.2) represents an
{\em exponential chirping} valid also for black holes.
For an elementary discussion of
Doppler shift for accelerated motion see Ref.[\refcite{coc}].
It is worthwile to mention
that in the semiclassical treatment of black hole physics one
is usually dealing with chirped signals, since the WKB functions are
generally of
variable wavelength, and by meeting supplementary conditions on their
derivatives they are made to look as much as possible like fixed
linear combinations of plane waves.
On the other hand, in the case of wave packets
one is always working with the average frequency of the
wave packets (see the second paper of Jacobson,\cite{jac} or the paper
of Frolov and Novikov on the dynamical origin of black hole entropy.\cite{fn})

The technique of producing plasma fronts/windows in a gas by laser breakdown,
and the associated frequency upshifting phenomena (there are also downshifts)
of the electromagnetic waves interacting with such windows
are well settled since about twenty years,
and blue shifts of about $10\%$ have been observed
in the transmitted laser photon energy.

In his paper, Yablonovitch works out a very simple model of a {\em linear}
chirping due to a refractive index linearly decreasing with time,
$n(t)=n_{0}-\dot n t$, implying a Doppler shift of the form
$\omega \rightarrow \omega[1+\frac{\dot n}{n} t]\sim \omega[1+\frac{a}{c} t]$.
To have accelerations $a =10^{20}g_{\oplus}$ the laser pulses should be
less than 1 picosecond.
Even more promising may be the nonadiabatic photoionization
of a semiconductor crystal in which case the refractive index
can be reduced from 3.5 to 0 on the timescale of the optical
pulse. As discussed by Yablonovitch, the pump laser has to be tuned
just below the Urbach tail of a direct-gap semiconductor in order
to create weakly bound virtual electron-hole pairs which contribute a large
reactive component to the photocurrent since they are readily polarized. The
background is due to the bremsstrahlung
 emission produced by real electron-hole pairs, and to diminish it
one needs a crystal with a big Urbach slope (the Urbach tail is an
exponential behavior of the absorption coefficient).

On the other hand, Yablonovitch remarked that
the experimental interpretation is highly
ambigous, since one may consider the phenomenon to be a single-cycle
microwave squeezing and/or an inverse quadratic electro-optic
effect with zero-point photons as input waves, and more
theoretically as Unruh effect and nonadiabatic Casimir effect.
Besides, one should notice the difference between the laboratory and
the black hole/hyperbolic chirping. The former is {\em linear}, whereas the
latter is {\em exponential}.

The `plasma window' of Yablonovitch was criticized in the important paper
by Dodonov,
Klimov, and Nikonov [DKN],\cite{dkn} on the grounds that we are not in the
case of exponentially small reflection coefficient as required to get
a Planck spectrum from vacuum fluctuations. At the general level,
one may argue that
nonstationary Casimir effects may produce some deformed Planck distributions,
and only in particular cases purely Planck distributions. As a matter of fact,
depending on the nonstationarity, one may obtain very peculiar photon spectra,
and this might be of great interest in applied physics.
DKN showed explicitly that an
exponential `plasma window', for which
presumably the modulation `depth' is the effective Unruh temperature, does
not produce a Planck spectrum. However, for a parametric function displaying
the symmetric Epstein profile one can get in the adiabatic limit a
`Wien's spectrum' with the effective temperature given by the logarithmic
derivative of the variable magnetic permeability with
respect to time.
According to
DKN this corresponds to a `dielectric window' and not to a `plasma window'.
The experimental
realization of nonstationary Casimir effects are either resonators with
moving walls, as first discussed by Moore,\cite{mo} or resonators with
time-dependent refractive media as discussed by DKN. On the lines
of Yablonovitch, Hizhnyakov,\cite{hiz} studied the sudden changes of the
refractive index caused by the excitations of a semiconductor near a
band-to-band transition in the infrared by a synchroneously pumped
subpicosecond dye laser, and also refered to the anology with Hawking
and Unruh effects. Very recently, C.K. Law,\cite{law} combined the moving
walls of Moore with the dielectric medium with time-varying permittivity
in a one-dimensional electromagnetic resonant cavity, obtaining an effective
quadratic Hamiltonian, which is always required when we want to discuss
nonstationary `particle production' effects.

Another interesting solid-state black hole emitting analog has been put
forth by Resnik,\cite{res} and refers to surfaces of singular electric
and magnetic permeabilities.

\vspace*{0.2cm}

\section{Unruh Effect and Channeling}

An Erevan 
group,\cite{dar} has proposed to measure the Unruh radiation
emitted in the Compton scattering of the channeled particles with
the Planck spectrum of the inertial crystal vacuum. The proposal
is based on the fact that the crystallographic fields are acting with
 large transverse accelerations on the channeled particles.

The estimated transverse proper acceleration for positrons
 channeled in the
(110) plane of a diamond crystal is $a = 10^{25}\gamma$ cm/s$^{2}$, and
at a $\gamma = 10^{8}$ one could reach
 $10^{33}$ cm/s$^{2}$ = $ 10^{30}g_{\oplus}$.

Working first in the particle instantaneous rest frame, the Erevan
group derived the spectral angular distribution of the Unruh photons
in that frame. By Lorentz transformation to the lab system they got the
number of Unruh photons per unit length of crystal and averaged over
the channeling diameter. At about $\gamma = 10^{8}$ the Unruh intensity,
i.e., the intensity per unit pathlength of the Compton scattering on the
Planck vacuum spectrum
becomes comparable with the Bethe-Heitler bremsstrahlung
($dN_{\gamma}/dE\propto 1/E$, and mean polar emission
angle $\theta =1/\gamma$).

Incidentally, there is a parallel with some experiments,\cite{deh,bin,cer}
 performed at LEP, where the scattering of the LEP beam from the
thermal photon background in the beam pipe has been measured (the
black body photons emitted by the walls of the pipe have a mean energy
of 0.07 eV). Fortunately the effect is too small to affect the lifetime
of the stored beams.

An eye should be kept open on such phenomena like parametric x-ray
production by relativistic particles in crystal,\cite{dub,cat}
as well as on other crystal- assisted phenomena.\cite{cry}

In another work of the armenian group,\cite{dar1}
the same type of calculations
was applied to estimating the Unruh radiation generated by TeV
electrons in a uniform magnetic field as well as in a laser field.
The Unruh radiation becomes predominant over the synchrotron
radiation only when $\gamma = 10^{9}$ for $H = 5\cdot 10^{7}Gs$ and
consequently it is impossible to detect it at the SLC.
Supercolliders with bunch structure capable of producing magnetic
fields of the order $10^{9} G$ are required.
Pulsar magnetospheres are good candidates for considering such a Unruh
radiation.

A circularly polarized laser field seems more promising since in this
case the Unruh radiation could be detectable at lower magnetic fields
and energies ($\gamma =10^{7}$). This is due to the fact that the proper
centripetal acceleration of the electron is
$a=2\omega\gamma\eta\sqrt{1+\eta ^{2}}$, where
$\omega $ is the frequency of the electromagnetic wave, and
$\eta =e\epsilon /m\omega$ ($\epsilon$ being the amplitude of the field).
\vspace*{0.2cm}

\section{ Hawking-like Effects and Free Electron Lasers (FELs)}

In principle, FELs might be a means to put into evidence
Unruh radiation as well as Hawking radiation.\cite{ro,ro1}

We first recall that in general relativity, it is well known the so-called
complexification trick/procedure,\cite{new}
which leads to new solutions of Einstein equations
from a given solution. In particular, A. Peres,\cite{per1}
 has shown long ago that by the
complexification of the isotropic Schwarzschild metric one could get a
gravitational tachyon, i.e., a super-light, extended ($r=2M$) gravitational
source,\cite{tac} (one may also call it a quasi-Minkowski metric with its
deviation from flatness of the form $f(Z-uT)$, with $u>1$, Maxwell constant is
unity). According to Peres
such procedures have been discussed by N. Rosen already in 1954.\cite{nro}
By this technique a closed horizon is changed
into an open (cone-like) horizon. The Mach cone and the Cherenkov shock
wave are common examples of open horizons. As far back as
1910, a paper of H. Bateman has the title
{\em ``Transformations of coordinates
which can be used to transform one physical problem into another''}.\cite{bat}
Jacobson and Kang,\cite{jk} have recently
investigated the conformal invariance of black hole temperature. They showed
that this is fulfilled for stationary black holes under those conformal
transformations being the identity at infinity.

On the other hand, the electromagnetic emission always makes an
important contribution to the radiation of a black hole horizon.\cite{p76}.
With this in mind we have to add to the complexification trick a second trick,
in which gravitation is supposed to be equivalent to an optical medium.
 This is an old but not very used method (Einstein was aware of it,
 and the initiators were W. Gordon, I. Tamm, and L.I. Mandelstam, who wrote
papers in the twenties). The interested reader may consult some more recent
literature.\cite{opt} It is as if in this case
 gravitation gets rid of its fundamental character turning into the
 constitutive equations of a dielectric medium with a variable
  refractive index. The constitutive relations of gravitational media are
$$
D_{i}=\epsilon _{ik}E_{k}-(\vec G \times \vec H)_{i}
\eqno(10.1)
$$
$$
B_{i}=\mu _{ik}H_{k}+(\vec G \times \vec E)_{i}
\eqno(10.2)
$$
where $\epsilon _{ik}=\mu _{ik}=-(g)^{1/2}g^{ik}/g_{00}$ and
$G_{i}=-g_{0i}/g_{00}$. The case $g_{0i}\neq 0$ is related to birefringence.
For FRW metrics one should use the following form of the
gravitational dielectric parameters
$$\epsilon _{ik}=\mu _{ik}= f(\rho)\delta _{ik}   \eqno(10.3)$$
The functional form of $f(\rho)$ depends on the cosmological and/or
black hole model one has in mind. For example in the de Sitter case
$f(\rho)\propto (1+\rho ^{2}/4R^{2})^{-1}$,
we see that the dielectric medium is just the
Maxwell fisheye lens. This is a spherical lens
with an index of refraction that can be written down in the form
$$
n(\rho) = {\frac{n_{0}}{1 + \alpha^{2}\rho^{2}}}
\eqno(10.4)
$$
for $\rho < R$. The constant $\alpha$ gives the constant optical gradient.
At the present time the GRIN (graded index) technology,\cite{moo}
is at the level
of $\alpha$ = 0.1-0.2 mm$^{-1}$. GRIN spheres were obtained for the first
time in 1986 by means of a modified suspension polymerization
technique.\cite{koi}

In our previous works,\cite{ro,ro1} we gave some hints
for studying the electromagnetic
radiation of the black hole horizon and the Unruh effect on the equivalent
scheme of a Cherenkov-Walsh FEL \cite{wa} with a GRIN lining.
  In such a FEL a relativistic electron beam of
 very good quality passes over a thin dielectric guide or through
 a channel in it, interacting with the axial component of the TM
 modes of the guiding structure. The stimulated emission occurs in
 the modes with a phase velocity slightly less than the velocity of the beam.

There might be a chance for Hawking-like effects to be seen in this
experimental configuration if and only if the lining structure is chosen to
be a GRIN material with a very high optical gradient (one may think
of quartz and fused silica which are common materials in GRIN optics).
Of course, the $\gamma$ of the beam should be very high. In this
setting, Hawking-like noise would be related to the waveguide dispersion
of the liner. Also gas-loaded FELs considered by Pantell's group should
be taken into account,\cite{pan} as well as plasma lining.
The estimated temperature of the Cherenkov wake in an inhomogeneous
lining material is
$$
T=\frac{\hbar c}{2\pi k}\frac{dn}{d\rho}
\eqno(10.5)
$$
The present-day optical gradients (0.2 mm$^{-1}$) could generate a
thermal effect of only 0.7 K.
Besides, there are many sources of noise in FEL devices (the most
common is the shot noise,\cite{ben}) and moreover, a lot of other phenomena
are waiting to be better understood before addressing more exotic and
minute effects. For example Sessler,\cite{ses} in his 1989 CAS
lecture ``Prospects for the FELs" speaks about an untowerd number of new
effects and discusses superradiance, plasma self-focusing, chirping, and
quantum mechanical behavior for electrons and photons in FEL settings, so
 clearly it will be very difficult to disentangle either `Hawking' or `Unruh'
 effect by means of FELs.

In addition, Becker and collaborators,\cite{bec} have commented
on testing the photon-photon sector of quantum electrodynamics (i.e.,
nonlinear effects in QED) with bright short-wavelength FELs with a high
 repetition rate.

Finally, different types of strophotron FEL configurations, which are based on
the channeling principle,
should be of interest for the Unruh radiation.\cite{og}

\vspace*{0.2cm}

\section{ Unruh Effect and Anomalous Doppler Effect (ADE)}

When studied with the detector method, the Unruh effect for a detector
with internal degrees of freedom is very close to the
anomalous Doppler effect (ADE), since in both cases the quantum detector
is radiating `photons' while passing onto the upper level and not on the
lower one.\cite{frol} It is worthwhile to note that the ADE-like concept
has been used by Unruh and Wald,\cite{wal} without referring to it
explicitly, when they have considered the Unruh effect for
a uniformly accelerated quantum detector looked upon from the inertial
reference frame. Their main and well-known conclusion was that emission
in an inertial frame corresponds to absorption from the Unruh's `heat bath' in
the accelerated frame. Essentially one may say the following.

(i) {\em For the observer placed in the noninertial frame the `photon'
is unobservable (it belongs to the left wedge in the Rindler case)}.

(ii) {\em When the observer places himself in an inertial reference frame,
he is able to observe both the excited quantum detector (
 furnishing at the same time energy to it) and the `photons'. By
writing down the energy-momentum conservation law he will be inclined
to say that the `photons' are emitted precisely when the detector is
excited}.

There is not much difference between the discussion of Unruh and Wald and
some Russian papers distributed over more than 40 years belonging to
Ginzburg and Frank. A quantum derivation of the formula for the Doppler
effect in a medium has been given by these authors already in 1947,\cite{47}
and more detailed discussion has been provided by Frank in
the seventies and eighties.\cite{798} See also the recent review paper
of Ginzburg.\cite{gi93}

Neglecting recoil, absorption, and dispersion (a completely ideal case)
the elementary radiation events for a
two-level detector with the change of the detector proper energy
denoted by $\delta\epsilon$ are classified according to the photon
energy formula,\cite{frol}
$$
\hbar\omega ={ -\frac{\delta\epsilon}{D\gamma}}
\eqno(11.1)
$$
where $\gamma$ is the relativistic velocity factor ($\gamma> 1$) and
D is the Doppler directivity factor
$$
D = 1 -( vn/c )\cos{\theta}
\eqno(11.2)
$$
The discussion of signs in Eq.(11.1) implies 3 cases as follows:

D$ >$ 0 for normal Doppler effect (NDE, $\delta\epsilon<0$)

D = 0 for Cherenkov effect (CE, $\delta\epsilon=0$, undetermined case)

D$ <$ 0 for anomalous Doppler effect (ADE, $\delta\epsilon>0$).

Consequently, for a quantum system endowed with internal degrees of
freedom the stationary population of levels is determined by the
probability of radiation in the ADE and NDE regions. The possibility
of doing population inversion by means of ADE has been tackled in the Russian
literature since long ago. A quantum system with many levels
propagating superluminally in a medium has been discussed for the
first time by Ginzburg and Fain in 1958.\cite{gf}
The inverse population of levels by means of ADE or a combination of
ADE and acceleration may be enhanced whenever the ADE region is made
larger than the NDE region. This is possible, e.g., in a medium with
a big index of refraction.
Naryshkina,\cite{nar} found already in 1962 that the radiation of
longitudinal waves in the ADE region is always greater than in the NDE
region, but apparently her work remained unnoticed until 1984, when
Nemtsov,\cite{nem} wrote a short note on the advantage of using ADE
longitudinal waves to invert a quantum system propagating in an isotropic
plasma.
The same year, Nemtsov and Eidman,\cite{nee} demonstrated inverse
population by ADE for the Landau levels of an electron beam propagating
in a medium to which a constant magnetic field is applied.
More recently, Kurian, Pirojenko and Frolov [KPF],\cite{kpf} have shown that
in certain conditions (for certain range of the parameters), a detector
moving with constant superluminous velocity on a circular trajectory
inside a medium may be inverted too. Bolotovsky and Bykov,\cite{bob}
have studied the space-time properties of ADE on the simple case of a
superluminous dipole propagating in uniform rectilinear motion in a
nondispersive medium. These authors are positive with the separate
observation of the ADE phenomenon for this case.

The radiation of a uniformly moving superluminal neutral polarizable
particle has been studied by Meyer.\cite{mey}
Frolov and Ginzburg,\cite{frol} remarked that this case is an
analog of ADE due to zero-point fluctuations of electric polarizability.

Moreover, we can modify the index of refraction in the Doppler factor
in such a manner as to get the ADE conditions already at sublight
velocities. In this way a more direct link to the Unruh effect is
available, as has been shown also by
 Brevik and Kolbenstvedt.\cite{bre} These authors studied in detail the
 DeWitt detector moving through a dielectric nondispersive medium with
 constant velocity as well as with constant acceleration, giving in
 first order perturbation theory formulas for transition probabilities
 and rates of emitted energy.

Let us mention here that one way to look at negative energy waves in
plasma physics is to consider them as a manifestation of induced ADE
elementary events discussed in the book of Nezlin.\cite{nez}
As a matter of fact, a number of authors have already dealt with the problem
of amplification and generation of electromagnetic waves based on ADE in
the field of quantum electronics.\cite{qe} For details on the nonlinear
instabilities in plasmas related to the existence of linear negative
energy perturbations expressed in terms of specific creation and
annihilation operators, and also for a discussion of the complete
solution of the three-oscillator case with Cherry-like nonlinear coupling,
one should consult the Trieste series of lectures delivered by
Pfirsch.\cite{pfi}
Also, Baryshevskii and Dubovskaya,\cite{bdu} considered ADE processes for
channeled positrons and electrons.
Moreover, Kandrup and O'Neill,\cite{kon} investigated the
hamiltonian structure of the Vlasov-Maxwell system in curved background
spacetime with ADM splitting into space plus time, showing the importance
of negative energy modes for time-independent equilibria.

\vspace*{0.2cm}

\section{Hawking-like Effects and Squeezing}  

Why is it that in the inertial vacuum we have only zero point
fluctuations but when changing to the coordinates of a noninertial
 reference frame, the new vacuum states, appropriately
 defined, are thermal-like states containing real photons ? Where do
 the real photons come from ?!

In our opinion, the most natural answer to such a paradox is given
in the context of squeezing. Any noninertial vacuum, no matter
how it is defined, is a squeezed vacuum with respect to the inertial
one. The squeezing parameter is related to the boost transformation
from inertial to noninertial coordinates. The point is that
squeezed vacuum states have a nonzero mean photon number
$$
<n> = \sinh^{2} r
\eqno(12.1)
$$
where r is the squeezing parameter characterizing the boost
transformation. Consequently, any noninertial/gravitational vacuum is
no longer a true vacuum, in the sense of having no real particles,
and the paradox is solved in a very convenient way.

The squeeze parametrization of the Bogolubov coefficients allows one
to accept the idea that some real photons show up in
the long quadrature of the squeezed zero point
fluctuations of a noninertial/gravitational vacuum. From this squeezing
perspective,
I do not favor the opinion of Barut and Dowling,\cite{bd} that
noninertial thermal baths do not contain real photons. Their claim
is that the photons are still bound to the body of the quantum noninertial
detector, though turned into dressed states. Of course, the relationship
between squeezing and dressed-state polarization (a variant of vacuum
polarization) is an interesting open problem for quantum physics
in general, to which the concept
of {\em decoherence} may have a substantial contribution.

I recall here that already in 1976 Hawking wrote down the Bogolubov
transformations for the
Schwarzschild black holes as follows,\cite{h76}
\[  \left\{ \begin{array}{ll}
      a^{(1)}_{\omega}=j_{\omega}\\
      a^{(3)}_{\omega}=(1-x_{b})^{-1/2}
  (h_{\omega} - x^{1/2}_{b}g^{\dag}_{\omega})\\
      a^{(4)} _{\omega}=(1-x_{b})^{-1/2}
   (g_{\omega}-x^{1/2}_{b} h^{\dag}_{\omega})
            \end{array}
    \right.   \]
where $x_{b}=\exp{(-8\pi GM\omega)}$ is the single Bogolubov parameter of the
problem. The $a$ operators are annihilation operators for modes having zero
Cauchy data on the past null infinity and a suitable time dependence on the
past horizon, whereas the right hand side annihilation and creation operators
correspond to a different basis of three orthogonal families taking into
account the fact that an observer at future null infinity can measure only
components of the modes outside the future horizon.
Grishchuk and Sidorov,\cite{gri} used the Bogolubov transformations obtained
by Hawking to show that the $in$ and $out$ states are related by a two-mode
squeeze operator with the squeezing parameter in each mode given by
$$
\tanh^{2}r = \exp{(-8\pi G M\omega)}
\eqno(12.2)
$$
Moreover, the two-mode SBH squeeze operator $S(r, \theta)$ has an EPR form,
{\em i.e.}, $S(r, \pi)$, with $r$ given by Eq. (12.2). This squeeze operator
is to
be applied to the $y$ and $w$ quasiparticles in Hawking's notation, i.e.,
those having zero Cauchy data on the past infinity and complementary zero
Cauchy data on the horizon in terms of the affine parameter.
Then one might think of the equivalence of the black hole spacetimes with
some nonlinear optical media in which parametric down conversion
phenomena have been put into evidence and represent an extremely active
research field. I am tempted to call just {\em spacetime squeezing}
the black hole squeezing, unless one think of it as the concept
used by Bialynicka-Birula team,\cite{bib} some time ago for the squeezing
due to the most general case of nonuniform and time
dependent linear electromagnetic media. As discussed in Section 10,
the {\em gravitational media} corresponding to the cosmological models
are usually gyrotropic, with equal permittivity and permeability tensors.
Other examples are the
{\em factorized media} of [DKN],\cite{dkn} i.e., media with space-time
factorized dielectric permittivity and magnetic permeability, which for the
time being have no gravitational or more common analogs.
Moreover, Yurke and Potasek,\cite{yp} have shown in the quantum optical
context that parametric
interactions resulting in the two-mode squeezing provide a mechanism
for thermalization whenever one is observing only one mode of a
two-mode squeezed vacuum. The generalization to the black hole case
is straightforward and provides a reasonable explanation for the
overwhelmingly discussed black-hole information paradox.
An equivalent of Eq. (12.1) for parametric processes is
$$
<n>= \sinh ^{2} (\Omega \kappa t/4)
\eqno(12.3)
$$
where $\Omega$ is the frequency of the resonant field with respect to which
the parametric processes are achieved, $\kappa$ is the `depth' of the
modulation, and $t$ is time. However, the electrodynamic particle production
processes in the laboratory involves weak nonlinear parametric phenomena, and
for the time being one can make
only a formal comparison with the powerful parametric processes required to
really put into evidence Hawking and/or Unruh effect.

A strong claim that laboratory squeezing in fibers is equivalent to
Unruh effect has been made by Grishchuk, Haus, and Bergman [GHB].\cite{ghb}
To accomplish laboratory optical squeezing one needs
to supress classical noise and phase match the vacuum wave with the
exciting source. These two conditions are very well satisfied by
working with fibers. However, one should be aware of the fact that the optical
and the material Schroedinger equations, despite
their similarities, have also some essential differences, as they apply to
different situations.

Highly interesting open issues are the connections among photodetection
theory, squeezed states, and accelerated detectors on the lines of
Klyshko.\cite{kly} A model
electron detector similar to the DeWitt monopole detector has been considered
some time ago by Cresser, who on its base developed a theory of electron
detection and photon-photoelectron correlations in two-photon
ionization.\cite{cres}

A paper of J.T. Wheeler,\cite{w} on the so-called
gravitationally squeezed light is also to be mentioned before ending
this section. Wheeler derived an estimate for the amplitude of the squeezing
in the case of a beam of coherent light propagating in a gravitational field.
For an earthbound experiment his formula is $A=8g_{\oplus}/5\omega c$, and so
$\omega =10^{15}$ Hz implies the minute figure $A=5\cdot 10^{-23}$.
Applying the same formula to the Unruh
effect, it might be possible to observe squeezing of photons emitted from
particle colliders with an amplitude less than one percent for the
same $\omega$
and accelerations of $10^{21}$ cm/s$^{2}$. In the case of black holes, J.T.
Wheeler asserted that squeezing may be used to tell how many times a given
photon had orbited the black hole close to the r = 3M limit.

It will also be of interest to look at the antibunching properties of
black hole radiation (a property of the fourth order correlation function),
whereas squeezing is a property of second order ones.\cite{gar}
Usually, the statistics of a beam is characterized
by the Mandel parameter $Q$. The $-1$ value of this parameter corresponds to
pure states.


\section{Unruh Effect and Hadron Physics}        

We would like to mention here one of the first applications
of Hawking-like effects, namely to explain the thermal spectrum in the
transverse energy of the produced particles observed in
 high-energy collisions. Salam and Strathdee,\cite{sal}
  have considered Hawking effect of Kerr-Newman black solitonic
solutions in strong gravity to be responsible for the $E_{T}$ thermal
spectrum. Hosoya,\cite{hos} (see also Ref.[\refcite{hor}])
applied moving mirror effects to the thermal gluon
production, and recently the armenian group,\cite{dar2}
estimated the contribution of Unruh effect to the soft photon production
by quarks, as entailed into the observed
anomalous low $p_{T}$ photons in $K^{+}p$ interections at $P = 70 \;GeV/c$.

The idea of relating the hadronic temperature to the Unruh effect is
rather old.\cite{bar} One way to introduce a
hadronic temperature is in terms of Lorentz-squeezed hadrons.\cite{hk}
Also, Dey {\em et al.},\cite{dey} related the Unruh temperature to the
observed departure from the Gottfried sum rule for the difference of the
proton and neutron structure functions in deep inelastic electron
scattering. In fact, some of these considerations are not far from the
way Nikishov and Ritus,\cite{nik} tackled the electromagnetic cases.

Other vivid pictures have to do with the relationship between the limiting
Hagedorn temperature/maximal acceleration
 and the Hawking temperature,\cite{par} the space-time
 duality symmetry, and the role played by strings in the last stages of
 black hole evaporation.\cite{ver}

\vspace*{0.2cm}

\section{Conclusions and Perspectives}         

I provided a heuristic survey of the various proposals made so far to
detect the class of thermal-like vacuum noises, commonly known as
Hawking effect and Unruh effect. The proposals enumerated herein suggest
a transition from a pure {\em gedanken phase} to a real {\em experimental one},
but it is fair to say that we are still far from those precise statements
required by the definite experimental action.
This research field is extremely
rich covering a large range of physical situations, and I tried
to touch upon its many facets from a global and rather pedestrian standpoint.
Of course, these effects, as measured on some analogs in
terrestrial laboratories, are extremely tiny. Nevertheless, the analogies
developed over the years showed that other fields of physics may have
a contribution to the better understanding of the two effects. Moreover,
as a corollary, those fields of physics enriched themselves with some
unconventional pictures. However, one should be always aware of the
ambiguity of interpreting the produced effects as a more direct consequence
of the employed experimental method rather than in terms of
sophisticated theoretical effects. In other words, the question of the most
natural interpretation is always the most stringent one when considering
analogies from the experimental point of view.

Perhaps one of the best applications of these conceptual effects is in the
areas of optical and electrodynamical
radiometry, since they clearly possess those universal qualities usually asked
for in those fields. My feeling at the end of the survey is that actually
the goal is not so much to try to measure a `Hawking' or a `Unruh' effect.
Being ideal concepts/paradigms, what we have to do in order to put them to
real work is to make them interfere with the many more `pedestrian'
viewpoints.

Another topic to be considered in more detail in the future is the connection
between Berry's phase and the noninertial/gravitational thermal-like
effects.\cite{han,chi,kug,hh} Indeed, Berry phase
can be related to the so-called Wigner angle (Lorentz transformations in
nonparallel directions do not commute involving a rotation angle) and also
to the Thomas
precession (measuring the time rate of Wigner rotations, and usually
associated to spin-orbit couplings,\cite{ma91}) which in turn could also be
considered in the class of squeezing phenomena. \cite{chi,hh}

Last but not least, the clear-cut aspects of Unruh effect in the realm
 of nonlinear (multiphoton) quantum electrodynamical effects,\cite{mcd} (the
 case of Hawking effect is similar within the squeezing perspective)
should be further studied taking into account the `quasi-feasibility' of
some proposed experimental schemes. As Prof. Keith McDonald,\cite{pc} recently
communicated to the
author, {\em it is useful to continue looking for new ways to
explore such effects}. At the same time, we shouldn't be
overenthusiastic about these highly ideal effects; the nonlinear physics
is extremely reach in all sorts of effects coming into play at some
curious length and time scales that might be assembled from various
combinations
of the coefficients in some nonlinear partial differential equations, that
usually enter the mathematical description of the complicated physical
processes that we were writing here about.

Anyway, the
correspondence between semiclassical electrodynamics and semiclassical
gravity within the pair creation regime should be further studied in order
to clarify their similarities and differences, and to appreciate better to
what extent a substantial amount of particle production might be well described
semiclassically. In this spirit, we draw attention to a recent paper
of Blencowe,\cite{blen}, who introduced and studied in some
detail an electrodynamical model, that one might call a `QED-Centauro'
phenomenon: an electrically neutral spherical object, entailing an equal
number densities of positive and negative charges exploding in such a way
that the negative charges leave the bubble as an expanding spherical shell.
Spontaneous pair creation analoguous to the Hawking effect occurs when the
potential energy difference between the shell and the core exceeds 2m$_{e}$,
where m$_{e}$ is the electron rest mass. Stephens,\cite{ste1} tried to
draw an
analogy between the one loop approximation of the pair production in a uniform
electric field and Hawking effect. Myhrvold,\cite{myh} commented on thermal
radiation from accelerated electrons.

As for the semiclassical gravity, one should notice the recent line of attack
suggested by Kuo and Ford,\cite{kf} and by Calzetta and Hu,\cite{ch}
in terms of a generalized
Langevin equation describing in Brownian manner the statistical behavior of
test particles moving in the fluctuating gravitational field. There are
several advantages of such an approach, among which a more transparent
interpretation of back reaction processes. It would be of interest to see
what will be the clarifying points brought in for Hawking effect in such a
picture.

There is lately a debate concerning the boundary conditions for the Unruh
effect \cite{rschool}. It points to a serious drawback of
all previous studies based only on
restricting the domain of definition of the fields by light cone (causal)
boundary conditions. It has been shown that the basic property of
hermiticity of the Hamiltonian of quantum field theories requires a particular
supplimentary boundary condition at the origin (i.e., the point in common
for the left and right Rindler wedges) which has not been
considered in deriving the `universal' Unruh effect giving the relationship
between the Rindler and Minkowski quantization schemes.
On the other hand, Grib \cite{grib} argued that in general
the light cone boundary conditions are not  of ``impenetrable wall" type.
The light cones are characteristic surfaces for wave equations and
causal conditions on them do not violate the wave equations.
Therefore a connection between the fields living in different regions of
the Minkovski space is possible. Of course, the matter of interpretation
of this connection is another deal.
Moreover, for the present author the behaviour of a noninertial
particle detector in empty Minkovski space is in many regards more
important and related indeed to real physics \cite{rreal}.



\nonumsection{Acknowledgements}
\noindent
This work was supported in part by the CONACyT Project 458100-5-25844E.


\nonumsection{References}


\end{document}